\documentclass[amsthm]{elsart}
\usepackage{amsfonts}
\usepackage{amsmath}
\usepackage{graphicx}

\begin{document}
\begin{frontmatter}
\title{Exact results in modeling planetary atmospheres--I. Gray atmospheres}
\runtitle{Exact results in planetary atmospheres modeling I}
\author [Paris]{L. Chevallier\thanksref{coresp},}
\author [Frankfurt] {J. Pelkowski,}
\author [Lyon] {B. Rutily}

\address [Paris]{Observatoire de Paris-Meudon, Laboratoire LUTH, 5 Place Jules Janssen, 92195 Meudon cedex, France}
\address [Frankfurt] {Institut f\"ur Meteorologie und Geophysik, J.W. Goethe Universit\"at Frankfurt, Robert Mayer Strasse 1, 
D-60325 Frankfurt a.M., Germany}
\address [Lyon]{Universit\'e de Lyon 1, Centre de Recherche Astronomique de Lyon, Observatoire de Lyon, 9 avenue Charles Andr\'e, F-69230 Saint-Genis-Laval, France ; CNRS, UMR 5574 ; Ecole Normale Sup\'erieure de Lyon, Lyon, France.}
\thanks[coresp]{Corresponding author. Tel.:+33-145-077-429; Fax.:+33-145-077-971.\\
\textit{Email address:} \texttt{loic.chevallier@obspm.fr} (L. Chevallier).}
\runauthor{Chevallier, Pelkowski and Rutily}
\begin{abstract}
An exact model is proposed for a gray, isotropically scattering planetary atmosphere in radiative equilibrium. 
The slab is illuminated on one side by a collimated beam and is bounded on the other side by an emitting and partially 
reflecting ground. We provide expressions for the incident and reflected fluxes on both boundary surfaces, as well as the temperature of the ground and the temperature distribution in the atmosphere, assuming the latter to be in local thermodynamic equilibrium. Tables and curves of the temperature distribution are included for various values of the optical thickness. Finally, semi-infinite atmospheres illuminated from the outside or by sources at infinity will be dealt with.   
\end{abstract}
\begin{keyword}
Planetary atmospheres; Radiative transfer; Isotropic scattering; Gray opacity; Radiative equilibrium; Temperature distribution.
\end{keyword}
\end{frontmatter}

\maketitle

\section{Introduction}

The modern phase in modeling radiation transfer through planetary atmospheres started with the seminal work of Humphreys \cite{humphreys1909}, and 
the contemporary elaborate attempt of Gold \cite{gold1909}, followed by the critical reappraisal of both attempts by Emden \cite{emden1913}. Of these 
works, Emden's is directly based on previous work by Schuster \cite{schuster1905} and Schwarzschild \cite{schwarzschild1906}, which, with Schwarzschild's second 
paper \cite{schwarzschild1914}, laid the foundation for the theory of radiation transfer. While those three papers may doubtlessly be regarded 
as ground-breaking, they necessarily remained rudimentary from our present vantage point in radiative transfer theory. It was not 
until the next decade that Milne \cite{milne1922} would deal with radiative energy transfer through planetary atmospheres in a 
mathematically rigorous manner. His important developments would not be superseded before the second half of the twentieth 
century, with the contributions by King \cite{king1963}, Ostriker \cite{ostriker1963}, Wildt \cite{wildt1966}, Schultis and Kaper \cite{schultisandkaper1969}, 
Stibbs \cite{stibbs1971} and Barkstrom \cite{barkstrom1974}, to name but those authors who attempted to model exactly the transfer through gray atmospheres.\\
By gray atmosphere we understand, as is customary, an atmosphere whose opacity does not depend on the frequency of the electromagnetic 
radiation involved. A semi-gray atmosphere is, on the other hand, one in which the opacity assumes just two different values 
in adjacent frequency intervals, conveniently distinguished as the ``visible'' and the ``infrared'' regions of the spectrum. 
It was Emden \cite{emden1913} who established this distinction, which, besides being more realistic, is susceptible to analytical treatment. 
The present paper is concerned with gray atmospheres in radiative equilibrium, while two further papers will be devoted to the case of 
semi-gray atmospheres.\\
In defining the model features, we shall assume the usual ingredients of the classical theory of planetary atmospheres to hold 
true: a plane-parallel atmosphere, subject to both illumination by a parallel beam of light impinging on one of its 
boundaries and to radiative heating through the other boundary, which is assumed contiguous to a partially 
reflecting and gray-emitting ground. Our purpose is to take the analytical developments as far as possible, under the assumption 
of a gray or semi-gray atmosphere in radiative equilibrium. We thereby wish to push the line of investigations recalled above 
farther, drawing from the theory of radiation transfer as it evolved in the last three decades. Our goal is to develop elementary but 
rigorous models which may be used not only to validate more realistic models based on numerical resources, but also to foster 
the understanding of radiative transfer in more comprehensive models of, say, the climates of planets, where it plays an 
important part in long-term evolutions.\\
The paper may be outlined as follows. Section 2 introduces the standard problem of planetary atmospheres. In Section 3, the fundamental integral 
equation underlying this problem is solved. Then follows the calculation of radiative flux densities at both boundary 
planes (Section 4), which enables the evaluation of the ground temperature (Section 5) and the atmospheric temperature (Section 6) of the planet. 
Next, in Section 7, we calculate the radiance of radiation leaving the atmosphere, as it represents the fundamental quantity in solving 
the inverse problem of the theory of planetary atmospheres. In the subsequent section, we deal with semi-infinite atmospheres heated either from 
the outside or from within by sources at infinity. Finally, we conclude the work with a summary, addressing at the same time some 
issues to be taken into account in forthcoming work.

\section{The problem stated}

We consider a plane-parallel, isotropically scattering atmosphere subject to the following two conditions:\newline
(\textit{i}) its opacity is gray, i.e., independent of frequency, and\newline
(\textit{ii}) radiative transfer throughout the atmosphere is to take place in local radiative equilibrium, which means that the flux density $F$ integrated over frequency is independent of position.\newline
To these defining physical assumptions we add the two following boundary conditions:\newline
(\textit{iii}) the atmosphere's upper surface is illuminated by a collimated solar beam (corresponding to a sun at infinity), carrying a steady frequency-integrated flux $F_0$ through a unit surface normal to the beam, while\newline
(\textit{iv}) the bottom boundary of the atmosphere is in contact with a gray, isotropically emitting and reflecting ground.\\
Note, however, that in Section 8, devoted to semi-infinite atmospheres, the boundary condition (\textit{iv}) will be replaced 
by an appropriate condition at infinity ({\it v}).\\ 
Turning to the basic equation of the model, let us begin by first introducing the frequency-integrated radiance field appropriate 
to our model assumptions. This function $I(\tau ,\mu,\varphi)$ depends on a positional and two angular arguments. 
The optical depth variable $\tau$ covers the range $[0,\tau_b]$, where $\tau_b$ is the optical thickness of 
the atmosphere. The top of the atmosphere is at $\tau=0$ and the bottom (i.e., the ground) at $\tau=\tau_b$. 
$\tau $ and $\tau_b$ are frequency-independent. $\mu$ stands for the direction cosine of the angle of 
incidence (zenith angle), referred to the outward normal at $\tau =0$, and $\varphi $ is the azimuth angle with respect to a 
chosen tangent direction in the plane $\tau = 0$. The radiance field obeys the following 
radiative transfer equation 
\begin{equation}
\label{eq1}
\mu\frac{\partial I}{\partial \tau }(\tau ,\mu,\varphi )=I(\tau ,\mu,\varphi)-S(\tau ),
\end{equation}%
wherein the source function $S$ is composed of the thermal emission and the radiation 
scattered isotropically by the elementary volumes of the atmospheric gas. The thermal emission may be written in the form $[1-a(\tau )]\mathfrak{B}^*(\tau )$, 
where $a(\tau )$ denotes the albedo for single scattering, which is independent of frequency by assumption, while $\mathfrak{B}^*(\tau)$ is an unknown 
function of $\tau$, deriving from the Planck function if the atmosphere is in local thermodynamic equilibrium. The scattering source function is expressed as $a(\tau )J(\tau )$, where $J(\tau )$ is the mean radiance, defined by Eq. (6) below. 
Hence 
\begin{equation}
\label{eq2}
S(\tau )=[1-a(\tau )]\mathfrak{B}^*(\tau )+a(\tau )J(\tau ).
\end{equation}%
The boundary condition on the (frequency-integrated) radiance at the upper surface $\tau =0$ is: 
\begin{equation}
\label{eq3}
I(0,\mu,\varphi )=F_{0}\delta (\mu+\mu_{0})\delta (\varphi -\varphi_{0})\;\;(-1\leq \mu<0<\mu_{0}\leq 1),
\end{equation}%
where the angular variables $(\mu_{0},\varphi _{0})$ define the direction of incident solar radiation, while $\delta$ 
denotes the Dirac distribution at 0. $\mu_0$ is positive because it refers to the zenith angle of the source (a point-like sun). At 
the bottom surface $\tau = \tau_b$ we require
\begin{equation}
\label{eq4}
I(\tau_b,\mu,\varphi )=\frac{1}{\pi }F^{\uparrow }(\tau_b)\qquad (0<\mu\leq 1),
\end{equation}%
the upward flux density at the bottom level $\tau = \tau_b$ being given by 
\begin{equation}
\label{eq5}
F^{\uparrow }(\tau_b)=F_{s}^{\uparrow}+r_{s}F^{\downarrow }(\tau_b).
\end{equation}%
The upward and downward fluxes are defined by Eqs. (9) and (10) below, respectively. $F_{s}^{\uparrow}$ refers to the flux radiated by the ground, to be 
calculated in Sec. 5, and $r_s$ is the Lambert flux reflectance of the ground (with its usual meaning, see e.g. \cite{thomasandstamnes1999}), which we assume to be given.\\ 
We next recall the definitions of mean radiance and net flux density 
\begin{equation}
\label{eq6}
J(\tau )=\frac{1}{4\pi }\int_{0}^{2\pi}\int_{-1}^{+1}I(\tau ,\mu,\varphi)\mathrm{d}\mu \mathrm{d}\varphi , 
\end{equation}%
\begin{equation}
\label{eq7}
F(\tau )=\int_{0}^{2\pi}\int_{-1}^{+1}I(\tau ,\mu,\varphi )\mu \mathrm{d}\mu \mathrm{d}\varphi ,
\end{equation}%
all quantities being integrated over frequency. Equation (7) may be rewritten in the form 
\begin{equation}
\label{eq8}
F=F^{\uparrow }(\tau )-F^{\downarrow }(\tau ),
\end{equation}%
in which the upward and downward flux densities are respectively given by
\begin{equation}
\label{eq9}
F^{\uparrow }(\tau )=\int_{0}^{2\pi}\int_{0}^{1}I(\tau,\mu,\varphi)\mu \mathrm{d}\mu \mathrm{d}\varphi \geq 0,
\end{equation}%
\begin{equation}
\label{eq10}
F^{\downarrow }(\tau )=-\int_{0}^{2\pi}\int_{-1}^{0}I(\tau ,\mu,\varphi)\mu \mathrm{d}\mu \mathrm{d}\varphi \geq 0.
\end{equation}
We recall that the condition of radiative equilibrium of a gray atmosphere means that the net flux density $F(\tau)$ 
is a constant $F$, which appears on the left-hand side of Eq. (8).\\
The formal solution of the radiative transfer equation (1) under the boundary conditions (3) and (4) reads
\begin{equation}
\label{eq11}
I(\tau,\mu,\varphi)=F_{0}\delta (\mu+\mu_{0})\delta (\varphi -\varphi_{0})\exp(\tau/\mu)-\frac{1}{\mu}\int_0^{\tau}S(\tau')\exp[(\tau-\tau')/\mu]\mathrm{d}\tau'
\end{equation}
for $-1\leq \mu < 0$, and
\begin{equation}
\label{eq12}
I(\tau,\mu,\varphi)=\frac{1}{\pi}F^{\uparrow}(\tau_b)\exp[-(\tau_b-\tau)/\mu]+\frac{1}{\mu}\int_{\tau}^{\tau_b}S(\tau')\exp[-(\tau'-\tau)/\mu]\mathrm{d}\tau'
\end{equation}
for $0<\mu\leq 1$. We may thus derive the following expressions for the mean radiance and the two flux densities previously defined, 
in terms of the source function: 
\begin{equation}
\label{eq13}
J(\tau )=J_{0}(\tau )+\frac{1}{2}\int_{0}^{\tau_b}E_{1}(|\tau -\tau'|)S(\tau')\mathrm{d}\tau',
\end{equation}%
\begin{equation}
\label{eq14}
F^{\uparrow }(\tau )=F_{0}^{\uparrow }(\tau)+2\pi \int_{\tau}^{\tau_b}E_{2}(\tau'-\tau)S(\tau')d\tau',
\end{equation}%
\begin{equation}
\label{eq15}
F^{\downarrow }(\tau )=F_{0}^{\downarrow }(\tau )+2\pi \int_{0}^{\tau}E_{2}(\tau-\tau')S(\tau')\mathrm{d}\tau',
\end{equation}%
where $J_{0}(\tau)$, $F_{0}^{\uparrow }(\tau )$ and $F_{0}^{\downarrow }(\tau)$ denote the corresponding contributions from the two boundary planes: 
\begin{equation}
\label{eq16}
J_{0}(\tau )=\frac{F_{0}}{4\pi}\exp(-\tau /\mu_{0})+\frac{1}{2\pi }F^{\uparrow }(\tau_b)E_{2}(\tau_b-\tau ),
\end{equation}%
\begin{equation}
\label{eq17}
F_{0}^{\uparrow}(\tau )=2F^{\uparrow }(\tau_b)E_{3}(\tau_b-\tau)\quad,\quad F_{0}^{\downarrow }(\tau )=\mu_{0}F_{0}\exp (-\tau /\mu_{0}).
\end{equation}%
The functions $E_{n}$ appearing on the right-hand sides of Eqs. (13)-(17) stand for the exponential integral functions of order 
$n\geq 1$, defined by
\begin{equation}
\label{eq18}
E_{n}(\tau )=\int_{0}^{1}\exp (-\tau /\mu)\mu^{n-2}\mathrm{d}\mu.
\end{equation}%
The main purpose of the present article is to calculate the incident and reflected fluxes at both boundary planes, 
as well as the temperature of the ground and the temperature profile of the atmosphere, in terms of the incident flux $F_0$ 
and the constant flux density $F$. Note that $F_0$ is truly a constant in our model, whereas $F$ depends on the position $\mu_0$ of the sun, as do 
all other quantities describing the radiation field. Beginning in the next section, this dependence will be made explicit.

\section{The fundamental integral equation solved}
The condition of radiative equilibrium entails that the atmosphere absorbs locally as much radiative energy as it emits, and thus
\begin{equation}
\label{eq19}
J(\tau,\mu_0)=S(\tau,\mu_0)=\mathfrak{B}^*(\tau,\mu_0)
\end{equation}%
for a gray atmosphere. The first equality is equivalent to the condition $F(\tau,\mu _{0})$ = $F(\mu_{0})$, as may be
seen from the integration of both members of the radiative transfer equation (1) over $\mu$ between -1 and +1. Replacing $J$ by $S$ on the left-hand side of Eq. (13), we obtain the following integral equation for the source
function: 
\begin{equation}
\label{eq20}
S(\tau,\mu_0)=J_{0}(\tau,\mu_0)+\frac{1}{2}\int_{0}^{\tau_b}E_{1}(|\tau -\tau'|)S(\tau',\mu_0)\mathrm{d}\tau'.
\end{equation}%
This equation shows that the frequency-integrated source function solves an integral equation of the Schwarzschild-Milne
type with albedo equal to 1. Physically, this is due to the fact that a gray atmosphere in radiative equilibrium is conservative for the frequency-integrated quantities describing the radiation field: there is neither loss nor gain of energy by absorption and emission, which is what the first equality in Eq. (19) is telling us. This fact by no means implies that the atmosphere's albedo for single scattering is everywhere equal to 1. Quite the contrary is the case when the second equality in (19) is imposed, which may be seen by inserting the first equality of Eq. (19) into the definition (11) for the integrated source function. One gets $[1-a(\tau)]S(\tau)=[1-a(\tau)]B^*(\tau)$,  leading to the second equality in (19) for layers $\tau$ in which $a(\tau)\neq 1$. Once the term $1-a(\tau)$ drops out, the albedo for single scattering of the atmosphere plays no role in the model, which means that it can take on
any value, and in particular be spatially variable.\\
In view of Eq. (16) and by the linearity of equation (20), its solution is found to be 
\begin{equation}
\label{eq21}
S(\tau,\mu_0)=\frac{1}{\pi}\,[\,\frac{F_{0}}{4}B(\tau ,\mu_{0})+F^{\uparrow}(\tau_b,\mu_0)\xi_{0}(\tau_b-\tau)\,]\,,
\end{equation}%
where the functions $B$ and $\xi_0$ solve the following integral equations:
\begin{eqnarray}
\label{eqs22/23}
B(\tau,\mu)&=&\exp(-\tau/\mu)+\frac{1}{2}\int_{0}^{\tau_b}E_{1}(|\tau -\tau'|)B(\tau',\mu)\mathrm{d}\tau',\\
\xi_{0}(\tau )&=&\frac{1}{2}E_{2}(\tau )+\frac{1}{2}\int_{0}^{\tau_b}E_{1}(|\tau -\tau'|)\xi_{0}(\tau')\mathrm{d}\tau'.
\end{eqnarray}%
The functions $B$ and $\xi_{0}$ are classical auxiliary functions of transfer theory, that have to be calculated for a slab of single scattering albedo equal to unity and an arbitrary optical thickness $\tau_{b}$. In what follows, their dependence on these two parameters will be implied. Thus, we shall write $B(\tau,\mu)$ instead of $B(1,\tau_{b},\tau ,\mu)$, $\xi_{0}(\tau)$ instead of $\xi_{0}(1,\tau_{b},\tau)$, and likewise for all the other auxiliary functions we shall introduce. As we just saw, the fact that these auxiliary functions are to be calculated for a conservative slab
(i.e., of unit albedo) is without relevance to the real value of the atmosphere's albedo, which may at any point be arbitrary.\\
The function $B$ was first introduced by Ambartsumian \cite{ambartsumian1942} in his attempt to 
solve the problem of diffuse reflection from a semi-infinite atmosphere. It was studied in detail by Busbridge \cite{busbridge1960} and coincides with van de Hulst's point-direction gain, which clarifies its physical meaning \cite{vandehulst1980}. An analytical expression for it in the case of a conservative atmosphere is given in the Appendix. The function $B$ satisfies the following relation referring to a finite slab of thickness $\tau_b$
\begin{equation}
\label{eq24}
B(\tau,\mu)=B(\tau_b-\tau, -\mu)\exp(-\tau_b/\mu),
\end{equation} 
and it reduces to the well-known $X$- and $Y$-functions on the two boundary layers:
\begin{equation}
\label{eq25}
B(0,\mu)=X(\mu)\quad,\quad B(\tau_b,\mu)=Y(\mu).
\end{equation} 
We refer to Chandrasekhar \cite{chandrasekhar1950} or Busbridge \cite{busbridge1960} for a survey of the principal properties of the functions $X$ and $Y$, 
along with their moments of order $n\geq 0$ 
\begin{equation}
\label{eq26}
\alpha_{n}=\int_{0}^{1}X(\mu)\mu^{n}\mathrm{d}\mu\quad,\quad \beta_{n}=\int_{0}^{1}Y(\mu)\mu^{n}\mathrm{d}\mu.
\end{equation}%
As to the function $\xi_0$, it may be seen from Eqs. (22)-(23) that it is proportional to the angular moment of order 0 of the function $B$, i.e.,
\begin{equation}
\label{eq27}
\xi_{0}(\tau )=\frac{1}{2}\int_{0}^{1}B(\tau ,\mu)\mathrm{d}\mu.
\end{equation}%
Equation (23) also shows that it can be interpreted as the escape probability function through the top surface of the atmosphere 
\cite{vandehulst1980}. It follows that
\begin{equation}
\label{eq28}
\xi_0(\tau)+\xi_0(\tau_b-\tau)=1, 
\end{equation}
which means that a photon, emitted at depth $\tau$ within a conservative finite slab, will eventually escape through either boundary surface. Setting 
$\tau=0$ in this relation and noting that the boundary values of the function $\xi_0$ are, 
according to Eqs. (25)-(27), given by
\begin{equation}
\label{eq29}
\xi_0(0) = \frac{1}{2}\,\alpha_0\quad,\quad \xi_0(\tau_b) = \frac{1}{2}\,\beta_0,
\end{equation}
one may readily deduce that the moments of order 0 of the functions $X$ and $Y$ fulfill the relation $\alpha_0+\beta_0=2$ in conservative 
media.\\
We refer to \cite{vandehulst1980} for graphs and tables of the function $g_0=2\xi_0$, and to \cite{rutilyetal2004} for the 
analytical calculation of the function $\xi_0$, including very accurate tables of this function. In the Appendix, we reproduce its expression for the conservative case.\\  
In the remainder, we will require the finite Laplace transform, at $1/\mu$ $(\mu>0)$, of the function $\tau \rightarrow B(\tau,\mu_0)$. 
Its expression
\begin{equation}
\label{eq30}
\int_0^{\tau_b}B(\tau,\mu_0)\exp(-\tau/\mu)\mathrm{d}\tau=4\mu\mu_0R(\mu,\mu_0)
\end{equation}  
can be found in \cite{busbridge1960}, p. 90. The reflection function $R(\mu,\mu_0)$ appearing on the right-hand side is 
the usual function describing the radiation reflected by a slab illuminated on its upper surface: see, e.g., Sec. 4.1.2 of van de Hulst \cite{vandehulst1980}. The latter reference produces the following expression for a
conservative slab (p. 194): 
\begin{equation}
\label{eq31}
R(\mu,\mu')=\frac{1}{4}\frac{X(\mu)X(\mu')-Y(\mu)Y(\mu')}{\mu+\mu'}.
\end{equation}
Integrating both members of (30) between 0 and 1 with respect to $\mu_0$, $\mu$ and both variables in turn leads to the following exact relations involving the functions $B$ and $\xi_0$:
\begin{equation}
\label{eq32}
\int_{0}^{\tau_b}\xi_0(\tau)\exp(-\tau/\mu)\mathrm{d}\tau=\mu R_1(\mu),
\end{equation}
\begin{equation}
\label{eq33}
\frac{1}{2}\int_0^{\tau_b}B(\tau,\mu_0)E_2(\tau)\mathrm{d}\tau=\mu_0R_1(\mu_0),
\end{equation}  
\begin{equation}
\label{eq34}
\int_0^{\tau_b}\xi_0(\tau)E_2(\tau)\mathrm{d}\tau=\frac{1}{2}\,R_{11},
\end{equation}  
where $R_1(\mu)$ and $R_{11}$ denote the moment and bimoment of order 1 of the reflection function, respectively, viz.
\begin{equation}
\label{eq35}
R_1(\mu)=2\int _{0}^{1}R(\mu,\mu')\mu'\mathrm{d}\mu'\quad,\quad R_{11}= 2\int_{0}^{1}R_1(\mu)\mu\mathrm{d}\mu.
\end{equation}
The calculation of these coefficients from expression (31) of the reflection function is well known: see, e.g., Display 9.1 on pp. 194-195 of 
\cite{vandehulst1980}. The result is, in the conservative case,
\begin{equation}
\label{eq36}
R_1(\mu)=1-\frac{1}{2}\beta_0(X+Y)(\mu)\quad, \quad R_{11}=1-\beta_0(\alpha_1+\beta_1).
\end{equation}
Now, by virtue of the properties (24) and (28) of the functions $B$ and $\xi_0$, the following relations may be derived from Eqs. (32)-(34):
\begin{equation}
\label{eq37}
\int_{0}^{\tau_b}\xi_0(\tau_b-\tau)\exp(-\tau/\mu)\mathrm{d}\tau=\mu [T_1(\mu)-\exp(-\tau_b/\mu)],
\end{equation}
\begin{equation}
\label{eq38}
\frac{1}{2}\int_0^{\tau_b}B(\tau_b-\tau,\mu_0)E_2(\tau)\mathrm{d}\tau=\mu_0[T_1(\mu_0)-\exp(-\tau_b/\mu_0)],
\end{equation}  
\begin{equation}
\label{eq39}
\int_0^{\tau_b}\xi_0(\tau_b-\tau)E_2(\tau)\mathrm{d}\tau=\frac{1}{2}\,[T_{11}-2E_3(\tau_b)].
\end{equation}  
Here $T_1(\mu)$ and $T_{11}$ are the moment and bimoment of order 1 of the transmission function of the atmosphere, whose expression is
\begin{equation}
\label{eq40}
T(\mu,\mu')=\delta(\mu'-\mu)\frac{\exp(-\tau_b /\mu)}{2\mu}+\frac{1}{4}\frac{Y(\mu)X(\mu')-X(\mu)Y(\mu')}{\mu-\mu'}
\end{equation}
[see Sec. 4.1.2 and Display 9.1, p. 194-195, of \cite{vandehulst1980}]. We thus have
\begin{equation}
\label{eq41}
T_1(\mu)=2\int _{0}^{1}T(\mu,\mu')\mu'\mathrm{d}\mu'\quad,\quad T_{11}= 2\int_{0}^{1}T_1(\mu)\mu\mathrm{d}\mu,
\end{equation}
and from Display 9.1 of \cite{vandehulst1980}
\begin{equation}
\label{eq42}
T_1(\mu)=\frac{1}{2}\beta_0(X+Y)(\mu)\quad, \quad T_{11}= \beta_0(\alpha_1+\beta_1).
\end{equation}
We see that indeed $R_1(\mu)+T_1(\mu)=1$ and $R_{11}+T_{11}=1$, as expected for a conservative atmosphere.\\
The solution (21) is, at this stage, merely formal, as it depends on the yet unknown flux density $F^{\uparrow}(\tau_b,\mu_0)$. 
In the following section, we shall calculate this flux by replacing the source function, given by expression (21), into Eqs. (14) and (15) for 
$\tau=0$ and $\tau=\tau_b$, respectively.

\section{Calculation of the flux density through the boundary planes}
According to Eqs. (14)-(15) and (17), the upward flux at $\tau=0$ and the downward flux at $\tau=\tau_b$ are respectively 
given by
\begin{eqnarray}
\label{eqs43/44}
F^{\uparrow}(0,\mu_0)&=&2F^{\uparrow}(\tau_b,\mu_0)E_3(\tau_b)+2\pi \int_{0}^{\tau_b}E_{2}(\tau)S(\tau,\mu_0)\mathrm{d}\tau,\\
F^{\downarrow}(\tau_b,\mu_0)&=&\mu_{0}F_{0}\exp (-\tau_b/\mu_{0})+2\pi \int_{0}^{\tau_b}E_{2}(\tau_b-\tau)S(\tau,\mu_0)\mathrm{d}\tau.
\end{eqnarray}%
Substituting expression (21) for $S(\tau,\mu_0)$ in these integrals and taking into account Eqs. (33)-(34) and (38)-(39), one obtains:
\begin{eqnarray}
\label{eq45/46}
F^{\uparrow}(0,\mu_0)&=&R_1(\mu_0)\mu_0 F_0+T_{11}F^{\uparrow}(\tau_b,\mu_0),\\
F^{\downarrow}(\tau_b,\mu_0)&=&T_1(\mu_0)\mu_0 F_0+R_{11}F^{\uparrow}(\tau_b,\mu_0).
\end{eqnarray}
The flux densities at the top of the atmosphere have values that follow immediately from the boundary condition (3) and flux constraint (8) at $\tau=0$:
\begin{equation}
\label{eq47}
F^{\downarrow}(0,\mu_0)=\mu_0 F_0\quad,\quad F^{\uparrow}(0,\mu_0)=\mu_0 F_0+F(\mu_0).
\end{equation}
We infer the two fluxes at the bottom boundary plane from Eqs. (45)-(47) 
\begin{eqnarray}
\label{eq48/49}
F^{\uparrow}(\tau_b,\mu_0)&=&\frac{1}{T_{11}}[\mu_0 F_0 T_1(\mu_0)+F(\mu_0)],\\
F^{\downarrow}(\tau_b,\mu_0)&=&\frac{1}{T_{11}}[\mu_0 F_0 T_1(\mu_0)+R_{11} F(\mu_0)].
\end{eqnarray}  
Note that $T_{11}\neq0$ if the atmosphere is finite ($\tau_b <\infty$), which is what we are supposing in this section.\\
These relations express the four boundary fluxes in terms of the incident flux $F_0$ and the flux constant $F(\mu_0)$. They have the virtue of simplicity 
when it comes to calculating the surface fluxes when at its top the atmosphere receives in a unit of time as much energy as it radiates [$F(\mu_0) = 0$]. 
At the same time, their physical interpretation is not as easily grasped as when the four boundary fluxes are expressed in terms of the incident flux $F_0$ 
and the flux emitted by the ground, $F_s^{\uparrow}(\mu_0)$. To see this, we express the constant flux $F(\mu_0)$ in terms of the fluxes $F_0$ and $F_s^{\uparrow}(\mu_0)$. 
But first we note that relation (8), for $\tau = \tau_b$, and relation (5) together imply 
\begin{equation}
\label{eq50}
F(\mu_0)=F_s^{\uparrow}(\mu_0)-(1-r_s)F^{\downarrow}(\tau_b,\mu_0),
\end{equation}
in which we now may substitute the expression (49), thus giving, after some rearrangements
\begin{equation}
\label{eq51}
F(\mu_0)=T^*F_s^{\uparrow}(\mu_0)-(1-r_s)T(\mu_0)\mu_0 F_0.
\end{equation}
$T(\mu_0)$ and $T^*$ denote respectively the flux transmittance and the spherical transmittance of the atmosphere, whose definitions will be recalled further on.\\
Replacing this expression for $F(\mu_0)$ in Eqs. (47)-(49), we find the following relations:
\begin{eqnarray}
\label{eq52/54}
F^{\uparrow}(0,\mu_0)&=&A(\mu_0)\mu_0F_0+T^*F_s^{\uparrow}(\mu_0),\\
F^{\uparrow}(\tau_b,\mu_0)&=&r_s T(\mu_0)\mu_0F_0+(T^*/\,T_{11})F_s^{\uparrow}(\mu_0),\\
F^{\downarrow}(\tau_b,\mu_0)&=&T(\mu_0)\mu_0F_0+T^*(R_{11}/\,T_{11})F_s^{\uparrow}(\mu_0),
\end{eqnarray}  
in which appear the flux reflectance (or plane albedo) $A(\mu_0)$ and the flux transmittance (or transmission) $T(\mu_0)$ of the atmosphere, whose expressions 
for a conservative atmosphere are \cite{sobolev1975,vandehulst1980}
\begin{eqnarray}
\label{eq55/56}
A(\mu_0)&=&1-(1-r_s)\frac{T_1(\mu_0)}{1-r_s R_{11}}=R_1(\mu_0)+r_s \frac{T_{11}}{1-r_s R_{11}}T_1(\mu_0),\\
T(\mu_0)&=&\frac{T_1(\mu_0)}{1-r_s R_{11}}.
\end{eqnarray}
The associated spherical quantities
\begin{equation}
\label{Eq57}
A^*=2\int_{0}^{1}A(\mu_0)\mu_0 \mathrm{d}\mu_0\quad,\quad T^*=2\int_{0}^{1}T(\mu_0)\mu_0 \mathrm{d}\mu_0
\end{equation}
are thus given by
\begin{eqnarray}
\label{Eq58/59}
A^*&=&1-(1-r_s)\frac{T_{11}}{1-r_s R_{11}}=R_{11}+r_s\frac{T_{11}^2}{1-r_s R_{11}},\\
T^*&=&\frac{T_{11}}{1-r_s R_{11}}.
\end{eqnarray}
Note also that the reflectances and transmittances are not independent of each other, as they verify 
\begin{equation}
\label{eq60}
A(\mu_0)+(1-r_s)T(\mu_0)=1\quad,\quad A^*+(1-r_s)T^*=1
\end{equation} 
in the case of a conservative atmosphere. Equation (51) clearly relates insolation of the planet, emission of its surface, and constant flux density $F(\mu_0)$ throughout its 
atmosphere. Notice that the latter is proportional to the coefficient $\beta_0$, which approaches zero when the optical thickness
$\tau_b$ tends to infinity. The constant flux density $F(\mu_0)$ necessarily vanishes for a semi-infinite atmosphere irradiated 
from the outside, which is to be formally confirmed in Subsection 8.2. Equations (52)-(54) extend the usual relations for atmospheres overlying an emissive ground. It will be 
noted that (53) reduces to $F^{\uparrow}(\tau_b,\mu_0)=F_s^{\uparrow}(\mu_0)$ when the ground is totally absorbing ($r_s=0$).\\
We introduce now what will be called the atmosphere's ``grayness factor'', defining it as the quotient between the flux density arriving at the ground and that reaching the top of the 
atmosphere
\begin{equation}
\label{eq61}
G(\mu_0)=\frac{F^{\downarrow}(\tau_b,\mu_0)}{\mu_0 F_0}.
\end{equation}
It may be rewritten in two different ways, depending on which expression is laid down, (49) or (54). The first one, 
\begin{equation}
\label{eq62}
G(\mu_0)=\frac{T_1(\mu_0)}{T_{11}}+\frac{R_{11}}{T_{11}}\frac{F(\mu_0)}{\mu_0F_0},
\end{equation} 
enables us to calculate the factor readily when the atmosphere receives as much energy as it radiates in a unit of time [$F(\mu_0)=0$], in which case 
\begin{equation}
\label{eq63}
G(\mu_0)=\frac{T_1(\mu_0)}{T_{11}} = \frac{1}{2}\frac{(X+Y)(\mu_0)}{\alpha_1+\beta_1}\,.
\end{equation}
This factor increases with $\mu_0$, as of course it should. It does not vanish for grazing incidence of the solar rays ($\mu_0=0$), but remains
 always smaller than 0.5. For normal incidence ($\mu_0 = 1$), whatever $\tau_b$, it is greater than unity, being enclosed within the limits 1 ($\tau_b=0$) 
and approximately 1.259, the value corresponding to $\tau_b=\infty$ (Subsection 8.2). For intermediate values of $\mu_0$, $G(\mu_0)$ may be smaller or greater 
than unity, depending on the values of $\tau_b$, as may be gleaned from published tables for the functions $X$, $Y$ and their first moments. Such tables may 
be found, for example, in \cite{sobouti1962}. The grayness factor is greatest for semi-infinite atmospheres irradiated from the zenith 
($\tau_b=\infty$, $\mu_0=1$).\\
The other expression for the grayness factor is obtained if in (61) we replace the flux reaching the ground by its expression (54), giving
\begin{equation}
\label{eq64}
G(\mu_0)=T(\mu_0)+T^*\frac{R_{11}}{T_{11}}\frac{F_s^{\uparrow}(\mu_0)}{\mu_0 F_0}.
\end{equation}
We see that this coefficient generalizes the transmission coefficient of the atmosphere, with which it is identical when the planet's 
surface does not radiate thermally. Moreover, the grayness factor provides an interesting alternative to the relation (51) linking the 
thermal emission of the ground to the flux constant $F(\mu_0)$, as may be seen at once from Eqs. (50) and (61):
\begin{equation}
\label{eq65}
F(\mu_0)=F_s^{\uparrow}(\mu_0)-(1-r_s)G(\mu_0)\mu_0 F_0.
\end{equation}
So as to be able to describe the global characteristics of the atmosphere, i.e., the ones at the planetary scale with a half-sphere irradiated by the sun, it would 
seem judicious to introduce a spherical grayness factor in the same way as the definition of the spherical transmission. The flux density of the radiation reaching the ground of the illuminated hemisphere 
is $2\pi R^2\int_0^1F^{\downarrow}(\tau_b,\mu_0)\mathrm{d}\mu_0$, where $R$ is the radius of the sphere. Dividing this flux by the solar flux $F_0\pi R^2$ 
intercepted by the sphere, we obtain the following definition of the spherical grayness factor:
\begin{equation}
\label{eq66}
G^*=\frac{2}{F_0}\int_0^1F^{\downarrow}(\tau_b,\mu_0)\mathrm{d}\mu_0=2\int_0^1G(\mu_0)\mu_0\mathrm{d}\mu_0.
\end{equation}
This average coefficient fulfills a number of interesting relations, obtained by integrating Eqs. (62) and (64) over $\mu_0$ between 0 and 1. Thus arises the globally averaged 
flux constant
\begin{equation}
\label{eq67}
F=\frac{1}{2}\int_0^1F(\mu_0)\mu_0\mathrm{d}\mu_0
\end{equation}
and likewise for the mean thermal flux emitted by the ground
\begin{equation}
\label{eq68}
F_s^{\uparrow}=\frac{1}{2}\int_0^1F_s^{\uparrow}(\mu_0)\mu_0\mathrm{d}\mu_0.
\end{equation}
The factor 1/2 in these expressions represents the ratio between the irradiated surface and the total surface of the atmosphere.\\
The formulas for the spherical grayness factor deriving from Eqs. (62) and (64) read:
\begin{equation}
\label{eq69}
G^* = 1+4\,\frac{R_{11}}{T_{11}}\,\frac{F}{F_0}=T^*[\,1+4\,\frac{R_{11}}{T_{11}}\,\frac{F_s^{\uparrow}}{F_0}\,],
\end{equation}
and by integrating (65) over $\mu_0$ one gets the relation
\begin{equation}
\label{eq70}
F=F_s^{\uparrow}-(1-r_s)G^*\frac{F_0}{4}.
\end{equation}
The first equation in (69) shows that the spherical grayness factor is equal to 1 when the atmosphere is in global radiative equilibrium, that 
is, when the global average flux $F$ vanishes. The second equality in (69) shows that the spherical grayness factor extends the concept of spherical 
transmissivity to the case of heating from below, both expressions being identical when there is no such heating by thermal radiation from the ground. 
Finally, an interesting variant of (70) arises from the elimination of the ratio $R_{11}/T_{11}$ between the two equations in (69), giving 
\begin{equation}
\label{eq71}
(G^*-1)T^*F_s^{\uparrow}=(G^*-T^*)F.
\end{equation}   
We shall take advantage of this result in the next section, devoted to calculating the temperature of the planet's surface.

\section{The ground temperature calculated}
To derive the temperature $T_s(\mu_0)$ of the ground from the flux $F_s^{\uparrow}(\mu_0)$ it radiates, we have to adopt 
an emission law for the ground. The simplest one is the gray law with (flux) emittance $e_s$, viz.
\begin{equation}
\label{eq72}
e_s \sigma T_s^4(\mu_0)=F_s^{\uparrow}(\mu_0),
\end{equation}
where $\sigma$ denotes the Stefan-Boltzmann constant. The flux $F_s^{\uparrow}(\mu_0)$ radiated by the ground may be evaluated from Eq. (65), thus enabling us to express the 
temperature of the ground in terms of the grayness factor: 
\begin{equation}
\label{eq73}
e_s \sigma T_s^4(\mu_0)=(1-r_s)G(\mu_0)\mu_0 F_0+F(\mu_0).
\end{equation}
This simplifies when the flux constant $F(\mu_0)$ vanishes, and the equilibrium relation $e_s+r_s=1$ is assumed to hold true, becoming
\begin{equation}
\label{eq74}
\sigma T_s^4(\mu_0)=G(\mu_0)\mu_0 F_0.
\end{equation}
Note that in this particular case the surface temperature is independent of the emitting and reflecting properties of the ground.\\
The temperature thus defined depends on the sun's elevation, which is to say that it assumes different values according to 
the location on the illuminated hemisphere. One may also define a global-mean surface temperature $T_s$ of a planet 
with radius $R$ in rapid rotation by equating the flux emitted by its surface, $4\pi R^2 e_s \sigma T_s^4$, to the flux irradiated by the surface 
of the illuminated hemisphere, $2\pi R^2\int_0^1 F_s^{\uparrow}(\mu_0)\mu_0\mathrm{d}\mu_0=2 \pi R^2 e_s \sigma \int_0^1T_s^4(\mu_0)\mu_0\mathrm{d}\mu_0$. Therefore,
\begin{equation}
\label{eq75}
\sigma T_s^4=\frac{1}{2} \int_0^1 \sigma T_s^4(\mu_0)\mu_0\mathrm{d}\mu_0.
\end{equation} 
It is thus easy to deduce from (73) and the first equality in (69) the following expression for the mean thermal flux 
emitted by the planet's surface:
\begin{equation}
\label{eq76}
e_s \sigma T_s^4=(1-r_s)\frac{F_0}{4}G^*+F=(1-r_s)\frac{F_0}{4}+\frac{F}{T^*}.
\end{equation} 
For an atmosphere in global radiative equilibrium ($F=0$), bounded by a surface for which $e_s+r_s=1$, the surface temperature 
is simply given by the relation
\begin{equation} 
\label{eq77}
\sigma T_s^4=\frac{F_0}{4},
\end{equation}
which shows that it depends only on the planet's distance from the sun.\\
As an illustrative numerical instance, let us determine the main global characteristics of the terrestrial atmosphere, assumed 
to be gray and in radiative equilibrium. It is known that the solar constant $F_0$ has a value of 1370 $\mathrm {W/m^2}$, and that the spherical 
albedo $A^*$ is close to 0.3. Adopting the mean value $r_s=0.095$ for the gray reflectivity of the Earth's surface ($r_s=0.15$ for visible light 
and $r_s=0.04$ for infrared radiation, see \cite{peixotoandoort1992}), the value of the coefficient of spherical transmission may be deduced from 
the stated value of the spherical albedo: $T^*=0.77$. We note in passing that the knowledge of $r_s$ and either $A^*$ or $T^*$ allows us to infer 
the optical thickness $\tau_b$ of the atmosphere: for that to be possible, it suffices to solve one of the equations $A^*(\tau_b, r_s)=0.3$ or 
$T^*(\tau_b, r_s)=0.77$ by bringing to bear (58) or (59). One finds $\tau_b=0.379$ as the optical thickness of the Earth's atmosphere when it is 
supposed to be both gray and in radiative equilibrium.\\
The mean surface temperature $T_s$ is observed to have a value around 288 K \cite{thomasandstamnes1999,peixotoandoort1992}, and  
the coefficient of emission of the surface is found to be $e_s=1-r_s = 0.905$. One may therefore deduce from the second equality in (76) the value 
of the mean flux constant $F$ = 33 $\mathrm{W/m^2}$, then from the first equality the value of the spherical grayness factor: $G^*=1.03$. The ratio 
$F/F_0$ is of the order of 2\%, which is close to a state of global radiative equilibrium. Let us remind the reader that strict global 
radiative equilibrium is incompatible with the equilibrium relation $e_s+r_s=1$ in our model, for it would imply relation (77), which is not 
satisfied by the chosen values for $F_0$ and $T_s$.  

\section{The atmospheric temperature calculated}
If now we assume the atmosphere to be in local thermodynamic equilibrium, its temperature structure may be determined by
substituting the expression (48) for the upward flux into the solution (21) for the source function $S(\tau,\mu_0)=\mathfrak{B}^*(\tau,\mu_0)=
(\sigma /\pi )T^{4}(\tau,\mu_0)$. Doing so, we obtain 
\begin{equation}
\label{eq78}
\sigma T^{4}(\tau,\mu_0)=\frac{F_{0}}{4}B(\tau,\mu_{0})+\frac{T_1(\mu_0)\mu_0 F_0 +F(\mu_0)}{T_{11}}\, \xi_0(\tau_b-\tau).
\end{equation}
This expression simplifies when $F(\mu_0)=0$. Dividing both members by $F_0$, defining the temperature $T_0$ by 
$\sigma T_0^4=F_0$, and recalling that $T_1(\mu_0)/T_{11}=G(\mu_0)$, we may express the temperature distribution within 
the atmosphere in the following dimensionless form:
\begin{equation}
\label{eq79}
\frac{T(\tau,\mu_0)}{T_0}=\left[ \frac{1}{4}\,B(\tau ,\mu_{0})+\,\mu_0 G(\mu_0) \xi_{0}(\tau_b-\tau)\right ]^{1/4}\,.
\end{equation}
The surface values of the atmospheric temperature may be calculated thanks to the surface values (25) and (29) of the functions $B$ and $\xi_0$. 
In this way, we get
\begin{equation}
\label{eq80}
\frac{T(0,\mu_{0})}{T_0}=\left[\frac{1}{4}X(\mu_{0})+\frac{1}{2}\beta_0\,\mu_0 G(\mu_0)\right]^{1/4}\,, 
\end{equation}
\begin{equation}
\label{eq81}
\frac{T(\tau_b,\mu_{0})}{T_0}=\left[\frac{1}{4}Y(\mu_{0})+\frac{1}{2}\alpha_0\,\mu_0 G(\mu_0)\right]^{1/4}\,.
\end{equation}
In particular, with the help of Eq. (74), one can write the temperature at the bottom of the atmosphere in the form
\begin{equation}
\label{eq82}
T^{4}(\tau_b,\mu_{0})=\frac{T_{0}^4}{4}Y(\mu_{0})+\frac{1}{2}\alpha_{0}\,T_{s}^{4}(\mu_{0})\,,
\end{equation}%
which clearly shows that it differs from the surface temperature $T_s(\mu_0)$. This discontinuity extends the well-known classical discontinuity occurring 
when $F_0=\sigma T_0^4 = 0$ \cite{thomasandstamnes1999,rutilyetal2004}. Since $Y(\mu_0)\geq 0$ and $(1/2)\alpha_0\leq 1$, the temperature of the atmosphere at the ground 
level may exceed or fall short of the ground temperature $T_s(\mu_{0})$, depending on the specific values of $\mu_0$ and $\tau_b$. Likewise, the following 
discussion will make clear that $T(0,\mu_{0})\leq T(\tau_b,\mu_{0})$ or $T(0,\mu_{0})\geq T(\tau_b,\mu_{0})$, depending on the value taken by the parameters $\mu_0$ and $\tau_b$.\\ 
The ratio $T(\tau,\mu_{0})/T_0$ is tabulated in Table 1 for $\mu_0 = 0.25$. This value has been chosen as appropriate to a global and annual average of the 
instantaneous  insolation of every point on a spherical planet revolving around the sun in a circular orbit. The ratio is plotted in Fig. 1 for 
$\mu_0 = 0.1, 0.25, 0.5$ and $1$. The selected values of $\tau_b$ are 0.01, 0.1, 1 and 10 for each one of these values of $\mu_0$. We made use of Eq. (A4) 
in the Appendix to evaluate the conservative function $\xi_{0}$. The present temperature profiles should of course be different from those 
of \cite{rutilyetal2004}, inasmuch as (79) is based on the assumption $F(\mu_{0})=0$.\\
\begin{table}
\caption{Values of $T(\tau,\mu_{0}) / T_0$ for $\mu_0=0.25$ and $\tau_b$ = 0.01, 0.1, 1 and 10.}
\label{tab1}
\centering
\begin{tabular}{ll}
\begin{tabular}[t]{lll}
\hline
\noalign{\smallskip}
$\tau_b$ & $\tau$ & $T(\tau,\mu_{0}) / T_0$ \\
\noalign{\smallskip}
\hline\noalign{\smallskip}
0.01  &	  0       &  0.7836 \\   
      &	  0.001   &  0.7836 \\
      &	  0.002   &  0.7835 \\
      &	  0.003   &  0.7834 \\
      &	  0.004   &  0.7833 \\
      &	  0.005   &  0.7830 \\
      &	  0.006   &  0.7828 \\
      &	  0.007   &  0.7826 \\
      &	  0.008   &  0.7823 \\
      &	  0.009   &  0.7821 \\
      &	  0.01    &  0.7817 \\
\noalign{\smallskip}
\hline\noalign{\smallskip}
1     &	  0     &  0.7886 \\   
      &	  0.1   &  0.7642 \\
      &	  0.2   &  0.7374 \\
      &	  0.3   &  0.7147 \\
      &	  0.4   &  0.6965 \\
      &	  0.5   &  0.6824 \\
      &	  0.6   &  0.6717 \\
      &	  0.7   &  0.6636 \\
      &	  0.8   &  0.6575 \\
      &	  0.9   &  0.6530 \\
      &	  1     &  0.6494 \\
\noalign{\smallskip}
\hline
\end{tabular}
&
\begin{tabular}[t]{lll}
\hline
\noalign{\smallskip}
$\tau_b$ & $\tau$ & $T(\tau,\mu_{0}) / T_0$ \\
\noalign{\smallskip}
\hline\noalign{\smallskip}
0.1   &	  0      &  0.7866  \\   
      &	  0.01   &  0.7857  \\
      &	  0.02   &  0.7835  \\
      &	  0.03   &  0.7809  \\
      &	  0.04   &  0.7780  \\
      &	  0.05   &  0.7751  \\
      &	  0.06   &  0.7720  \\
      &	  0.07   &  0.7688  \\
      &	  0.08   &  0.7655  \\
      &	  0.09   &  0.7621  \\
      &	  0.1    &  0.7583  \\
\noalign{\smallskip}
\hline\noalign{\smallskip}
10    &	  0   &  0.7886  \\   
      &	  1   &  0.6501 \\
      &	  2   &  0.6408 \\
      &	  3   &  0.6399 \\
      &	  4   &  0.6398 \\
      &	  5   &  0.6397 \\
      &	  6   &  0.6397 \\
      &	  7   &  0.6397 \\
      &	  8   &  0.6397 \\
      &	  9   &  0.6397 \\
      &	  10  &  0.6397 \\
\noalign{\smallskip}
\hline
\end{tabular}
\\
\end{tabular}
\end{table}
\begin{figure*}[htb]
\begin{center}
\begin{tabular}{cc}
\resizebox{0.47\hsize}{!}{\includegraphics{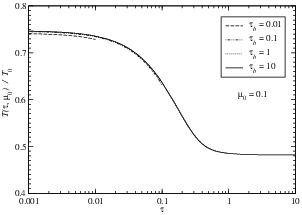}}
&\resizebox{0.47\hsize}{!}{\includegraphics{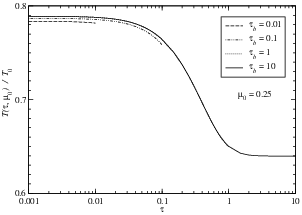}} \\
\resizebox{0.47\hsize}{!}{\includegraphics{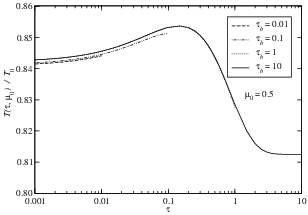}}
&\resizebox{0.47\hsize}{!}{\includegraphics{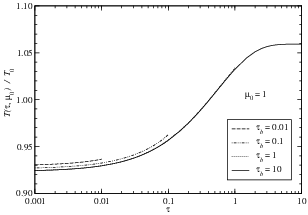}}
\end{tabular}
\end{center}
\caption{$T(\tau, \mu_{0}) / T_0$ versus $\tau$ for $\mu_0 = 0.1, 0.25, 0.5, 1$ and $\tau_b = 0.01, 0.1, 1, 10$. The curves for $\tau_b=1$ are 
not visible since they nearly coincide with that for $\tau_b=10$ over the range $[0, 1]$.}
\label{fig1}
\end{figure*}

The shape of the temperature curves shown in Fig. 1 depends on the angle of inclination of the solar rays. For $\mu_0=0.1$ 
and $\mu_0=0.25$, the temperature tends to decrease from a higher level-value in the surface layers of the atmosphere to a lower level-value 
in layers where the atmosphere is optically thick enough, which may be assumed to be the case when $\tau_b\geq1$. The temperature levels off 
roughly for $\tau\geq1$ or 2, and the atmosphere becomes virtually isothermal at greater optical depths. The two ``plateau'' temperatures 
are given by the expressions (80)-(81), and depend but weakly on the optical thickness $\tau_b$ of the atmosphere.\\
For greater values of $\mu_0$, $\mu_0=0.5$ and $\mu_0=1$ for instance, the temperature starts off {\it increasing} as we get into the atmosphere, 
a fact that reflects the greater atmospheric heating efficiency of the sun's radiation at solar incidence angles closer to the surface normal. 
The temperature reaches a maximum value when $\mu_0 = 0.5$, but for normal incidence ($\mu_0=1$) it does not cease to increase with depth, whatever 
the value of $\tau_b$. The fact that $T(\tau_b,\mu_{0})$ may be greater than $T(0,\mu_{0})$ for normal or nearly normal incidence of the solar rays may readily be 
understood by noting that these two extreme temperatures are close to their values in a semi-infinite atmosphere as we plunge deep enough into the latter, 
say for $\tau_b \geq1$. Indeed, from Eqs. (108) below, we have $T(\tau_b,\mu_{0})/T(0,\mu_{0})\sim T(\infty,\mu_{0})/T(0,\mu_{0})=(\sqrt{3}\mu_0)^{1/4}$, and this ratio exceeds unity whenever $\mu_0 \geq 1/\sqrt{3}$.
It is also under these circumstances that the grayness factor is necessarily greater than unity.\\
To illuminate further the behavior of $T(\tau,\mu_{0})$ according to the values of $\mu_0$ and $\tau_b$, we need to ascertain the temperature derivative 
at every point of the atmosphere, which calls for the previous determination of the derivatives of the functions $\tau \rightarrow B(\tau,\mu_0)$ and 
$\tau \rightarrow \xi_0(\tau_b-\tau)$, as can be seen from (79). Fortunately, these two derivatives may be expressed analytically in terms of a 
new auxiliary function, denoted by $\phi$, a function that has been thoroughly studied by Sobolev \cite{sobolev1963,sobolev1975}. While the expression for the derivative of $B$ is 
classic (see \cite{busbridge1960}, p. 91), the derivative of $\xi_0$ results from it by a simple integration over the angular variable $\mu$. We shall 
dispense with these calculations for finite atmospheres and instead carry them through only in the case of semi-infinite atmospheres (Subsection 8.2). We point out that 
a thorough study of the temperature variation in the neighborhood of both boundary planes would reveal that the derivatives of the temperature 
are {\it infinite} at $\tau=0$ and $\tau=\tau_b$, whereas the curves of Fig. 1 seem to suggest that they vanish, an illusion due to the chosen logarithmic 
scale for the abscissa.\\
We now define the global mean temperature $T(\tau)$ of a planet's atmosphere in fast rotation in much the same way as we 
defined its surface temperature, that is, through the relation
\begin{equation}
\label{eq83}
\sigma T^4(\tau)=\frac{1}{2}\int_0^1 \sigma T^4(\tau,\mu_0)\mathrm{d}\mu_0.
\end{equation}    
This temperature profile verifies the equation obtained by integrating both members of (78) over $\mu_0$, viz.
\begin{equation}
\label{eq84}
\sigma T^4(\tau)=\frac{F_0}{4}+\frac{F}{T_{11}}\,\xi_0(\tau_b-\tau).
\end{equation}
In an atmosphere in global radiative equilibrium, the average flux constant $F$ vanishes and the preceding relation shows that the averaged 
atmospheric temperature is constant and equal to $F_0/4$. This is just the value encountered for the surface temperature when one assumes both 
the equilibrium relation $e_s+r_s=1$ and $F=0$. Both assumptions thus imply that the ground and atmosphere are, on a global average, thermally uniform.  

\section{The radiance of the emerging radiation from the atmosphere}
To be able to calculate the radiance of the radiation field in the atmosphere, we need only substitute the expression (21) for the source function 
into the formal solution (11)-(12) of the radiative transfer equation, and then integrate analytically with respect to $\tau'$. Although 
such an integration is possible for finite media, it is rather tedious, due to the complexity of the functions $B$ and $\xi_0$ in the finite case. It is 
much more tractable for semi-infinite atmospheres, for which we shall give the result in the next section. Here, we limit ourselves to recalling 
the expression for the emergent radiance, a quantity useful in the solution of inverse problems raised by the theory of planetary atmospheres.\\
Inserting the expression (21) for the source function into relation (12) for $\tau=0$, one can perform the integration with respect to $\tau'$ thanks 
to Eqs. (30) and (37). We get for $0 < \mu \leq 1$:  
\begin{equation}
\label{eq85}
I(0,\mu,\mu_0)=\frac{1}{\pi}\left[\mu_0 F_0 R(\mu,\mu_0)+F^{\uparrow}(\tau_b,\mu_{0})(1/2)\beta_0(X+Y)(\mu)\right].
\end{equation}
We recall that the reflection function $R(\mu,\mu_0)$ is expressed by relation (31) in terms of the $X$- and $Y$-functions. It is clear that we can 
infer the flux $F^{\uparrow}(\tau_b,\mu_{0})$ radiated from the ground by measuring the radiation leaving the atmosphere, 
provided we know $\tau_b$, $F_0$ and $\mu_0$. Likewise, the ground temperature may be deduced from $F^{\uparrow}(\tau_b,\mu_{0})$ with the help of Eqs. (72), 
(5) and (46), should the reflectance and emittance of the ground be known.

\section{Semi-infinite planetary atmospheres}
We may distinguish two categories of semi-infinite atmospheres: those that are heated from the outside and those with internal sources at infinity, 
as in the case of stellar atmospheres. Of course, that does not prevent atmospheres from having both categories of sources. In this event, the radiative quantities 
required for their description are the sum total of quantities resulting from each category treated separately. That we may proceed in this manner is a 
consequence of the linearity of the transfer equation under the given boundary conditions.\\
As gray semi-infinite models are relatively simple, we shall be able to evaluate the radiance at each point of the atmosphere, a task we have laid aside 
in the finite case. To that end, new auxiliary functions will be required, whose analytical expressions we recall in what follows.
\subsection{The functions $H$, $\phi$ and $\eta$} 
The function $H$, first introduced by Hopf \cite{hopf1934} for conservative media, has been well studied and it is discussed 
in most treatises on radiation transfer, beginning with Chandrasekhar's and Busbridge's monographs \cite{chandrasekhar1950,busbridge1960}. 
In a conservative atmosphere, the function $H$ may be evaluated with the aid 
of the following exact formula, resulting from Mullikin's expression when the albedo approaches unity \cite{mullikin1964}:
\begin{equation}
\label{eq86}
H (\mu) = (1+\mu)\exp{\left[\mu \int_{0}^{1} \theta (v)\frac{\mathrm{d}v}{v(v+\mu)}\right]} \qquad (\mu \geq 0).
\end{equation}
It involves the functions defined for $0\leq v < 1$ by
\begin{equation}
\label{eq87}
\theta (v) = \frac{1}{\pi} \arctan \left[ \frac{\pi}{2} \frac{v}{T(v)} \right],
\end{equation}
\begin{equation}
\label{eq88}
T(v) = 1 - \frac{1}{2} v \ln \left(\frac{1+v}{1-v} \right).
\end{equation}
Continuous values on $[0,\pi]$ of the arctan function are used in the definition (87) of the $\theta$-function, i.e., the 
branch is not the principal one. The resulting function $\arctan(x/y)$ is usually denoted by $\mathrm{ATAN2}(x,y)$. With this 
choice, the function $v \to \theta(v)$ is continuous from $[0,1[$ to $[0,1[$, although $T(v)$ vanishes once in the interval 
$[0,1[$.\\
The function $\phi$, or resolvent function, is the other auxiliary function that has been widely studied, notably by Sobolev \cite{sobolev1963,sobolev1975}. For conservative 
media, we may calculate it by means of Minin's expression \cite{minin1958}
\begin{equation}
\label{eq89}
\phi(\tau)=\sqrt{3}+\frac{1}{2}\int_{0}^{1}\frac{g(v)}{H(v)}\exp(-\tau/v)\frac{\mathrm{d}v}{v} \quad(\tau>0),
\end{equation}
where
\begin{equation}
\label{eq90}
g(v) = \frac{1}{T^{2}(v) + [(\pi/2)v]^{2}} \qquad (0 \leq v < 1).
\end{equation}
The function $\eta$, on the other hand, is far less well-known than the preceding two functions, and only occasionally it has cropped up in the literature, 
those occasions having already been pointed out in Section 4 of \cite{bergeatandrutily1998}. We shall calculate it here for conservative media by letting 
$a \rightarrow 1$ in Eqs. (36) and (38) of that article, which yields for $\tau\geq 0$ and $\mu \in \mathbb{R}$:
\begin{equation}
\label{eq91}
\eta(\tau,\mu)=\mathrm{Y}(\mu)\frac{c(\mu)}{H(\mu)}\frac{\exp(-\tau/\mu)}{\mu}+\mathfrak{F}(\tau,\mu).
\end{equation}
Here Y($\mu$) denotes Heaviside's function, equal to 0 for $\mu<0$ and to 1 for $\mu>0$, and  
\begin{eqnarray}
\label{eq92}
\lefteqn{c(\mu)  =  \left\lbrace 
\begin{array}{ll}\displaystyle 1/T(\mu) & \rm \; if \;\; \mu \in ]-\infty,0[\cup]1,+\infty[,\\ 
\displaystyle g(\mu)T(\mu) & \rm \;if\;\;\mu \in[0,1[, \\
\displaystyle 0 & \rm \;if \;\;\mu =1, 
\end{array} \right.}
\end{eqnarray}
and
\begin{equation}
\label{eq93}
\mathfrak{F}(\tau,\mu)=\sqrt{3}+\frac{1}{2}\int_0^1\frac{g(v)}{H(v)}\exp(-\tau/v)\frac{\mathrm{d}v}{v-\mu}.
\end{equation}
The integral on the right-hand side has to be evaluated as a Cauchy principal value for $0<\mu<1$. For $\tau=0$ and $\mu\neq0$, this integral may 
be carried out analytically, whereby we find
\begin{equation}
\label{eq94}
\mathfrak{F}(0,\mu)=\frac{1}{\mu}[1-\frac{c(\mu)}{H(\mu)}].
\end{equation}
Consequently,
\begin{equation}
\label{eq95}
\eta(0,\mu)=\frac{1}{\mu}[1-\mathrm{Y}(-\mu)H(-\mu)].
\end{equation} 
Equations (91)-(93) also show that the function $\eta$ is bounded as $\tau \to \infty$, with limit equal to $\sqrt{3}$. Finally, these equations
 can be extended by continuity at $\mu=0$, and we retrieve the function $\phi$
\begin{equation}
\label{eq96}
\eta(\tau,0)=\phi(\tau)\quad(\tau > 0).
\end{equation}
The functions $H$, $\phi$ and $\eta$ are the building stones that enable us to construct almost all auxiliary functions needed in solving the transfer 
equation in a half-space, the function $B$ being one of them. Indeed, the latter is related to the functions $H$ and $\eta$ by means of the following relationship, valid 
for $\mu>0$ \cite{bergeatandrutily1998}
\begin{equation}
\label{eq97}
B(\tau,\mu)= H(\mu)\;\mu\;\eta(\tau,\mu).
\end{equation}   

\subsection{Semi-infinite atmospheres irradiated from the outside}
In this model category, to which, as a particular instance, belongs the atmosphere of Venus, the third condition $(iii)$ of Section 2 applies unchanged:\\
$(iii)$ At the upper surface of the atmosphere, the incident radiation is given by
\begin{equation}
\label{eq98}
I(0,\mu,\varphi, \mu_0,\varphi_0)=F_{0}\delta (\mu+\mu_{0})\delta (\varphi -\varphi_{0})\;\;(-1\leq \mu<0<\mu_{0}\leq 1),
\end{equation}%
while condition $(iv)$ is to be replaced by the following condition at infinity:\\
$(v)$ The radiance remains bounded as the depth in the atmosphere tends to infinity
\begin{equation}
\label{eq99}
\lim_{\tau\to \infty}I(\tau,\mu,\varphi,\mu_0,\varphi_0)<\infty.
\end{equation}
The main results for this model may be obtained from those derived for finite atmospheres by letting $\tau_b \to \infty$. All 
terms in Eqs. (12), (16) and (17) containing $F^{\uparrow}(\tau_b)$ vanish, because they are multiplied by a factor that tends 
to 0 as $\tau_b\to \infty$. Equation (21) thus becomes, in the semi-infinite case,
\begin{equation}
\label{eq100}
S(\tau,\mu_0)=\frac{F_{0}}{4\pi }B(\tau,\mu_{0}),
\end{equation}%
the function $B$ for conservative semi-infinite media being given by (97).\\
It is now possible to deduce the radiance at each point of the atmosphere. To that end, it suffices to substitute the present expression for the source 
function in the formal solution (11)-(12) of the transfer equation. The analytical integration over $\tau'$ then yields
\begin{eqnarray}
\label{eq101}
\lefteqn{I(\tau,\mu,\varphi,\mu_0,\varphi_0)=\mathrm{Y}(-\mu)F_{0}\delta (\mu+\mu_{0})\delta (\varphi -\varphi_{0})\exp(\tau/\mu)\nonumber}\\
&&\qquad\qquad\qquad\quad +\frac{1}{4\pi}\;H(\mu_0)\frac{\mu_0\eta(\tau,\mu_0)+\mu\eta(\tau,-\mu)}{\mu+\mu_0}\mu_0F_0 .
\end{eqnarray}
This result has been known since the end of the fifties, as reported by Danielyan in a commentary to his Eqs. (5)-(6) in \cite{danielyan1983}. 
At the upper boundary $\tau=0$, by (95) we retrieve the boundary condition (98) for $-1\leq \mu<0$, and also 
\begin{equation}
\label{eq102}
I(0,\mu,\mu_0)=\frac{1}{4\pi}\frac{H(\mu)H(\mu_0)}{\mu+\mu_0}\mu_0F_0 
\end{equation}  
for $0<\mu\leq1$. This is the well-known law of diffuse reflection for semi-infinite media \cite{chandrasekhar1950}, already retrieved by Stibbs \cite{stibbs1971} in 
the context of planetary atmospheres. It arises likewise from the finite-atmosphere counterpart (85) in the limit of infinite optical 
thickness $\tau_b$, since the functions $X$ and $Y$ tend to $H$ and 0, respectively, as $\tau_b\to\infty$. As in the finite case, we might have 
established it directly, without resorting to the expression for the radiation field within the atmosphere.\\
Deep within the atmosphere, the radiative field becomes, as indeed it should, asymptotically isotropic, with the limiting value obtained by letting $\tau \to \infty$ 
in (101) and by taking into account the fact that then $\eta(\tau,\mu) \to \sqrt{3}$. Consequently,
\begin{equation}
\label{eq103}
\lim_{\tau \to \infty}I(\tau,\mu,\varphi,\mu_0,\varphi_0)=\frac{\sqrt{3}}{4\pi}H(\mu_0)\mu_0F_0 .
\end{equation}
An isotropic radiance at infinity goes along with a vanishing net flux density at infinity, implying a vanishing net flux everywhere since the latter 
is constant. As a result, the upward and downward fluxes coincide everywhere, and their common value may be deduced from the expression (101) for the radiance. 
We get 
\begin{eqnarray}
\label{eq104}
\lefteqn{F^{\uparrow}(\tau,\mu_0)=F^{\downarrow}(\tau,\mu_0)=\frac{1}{2}\mu_0F_0H(\mu_0)\nonumber}\\
&&\qquad\qquad\qquad\qquad\qquad\times\int_0^1[\mu_0\eta(\tau,\mu_0)+\mu\eta(\tau,-\mu)]\frac{\mu\mathrm{d}\mu}{\mu+\mu_0}.
\end{eqnarray}
For $\tau=0$, the expression in brackets equals $H(\mu)$ by Eq. (95), and the so-called ``alternative form'' of the $H$-equation, 
as found for instance in \cite{vandehulst1980}, p. 162, shows that the integral is equal to $2/H(\mu_0)$. One thus definitely 
retrieves $F^{\uparrow}(0,\mu_0)= F^{\downarrow}(0,\mu_0) = \mu_0 F_0$. In the deep layers of the atmosphere, as the function $\eta$ tends to $\sqrt{3}$, both fluxes 
approach the following limiting value:
\begin{equation}
\label{eq105}
\lim_{\tau\to\infty}F^{\uparrow}(\tau,\mu_0)=\lim_{\tau\to\infty}F^{\downarrow}(\tau,\mu_0)=\frac{\sqrt{3}}{4} H(\mu_0)\mu_0 F_0.
\end{equation}
The grayness factor of a semi-infinite atmosphere, defined as $F^{\downarrow}(\infty,\mu_0)/\mu_0 F_0$ by analogy with (61), is thus given by
\begin{equation}
\label{eq106}
G(\mu_0)=\frac{\sqrt{3}}{4}H(\mu_0).
\end{equation}
There exist numerous tables of the function $H$ in the literature, for instance in the monograph by Chandrasekhar \cite{chandrasekhar1950}. Our own tables 
reveal that $G(\mu_0)$ exceeds unity for $\mu_0 > 0.665$, the maximum value 1.259 being reached for $\mu_0=1$. It may be noted that the expression (63) for the grayness 
factor of finite media does reduce to (106) as $\tau_b\to\infty$, since in that case $\alpha_1$ tends to the first-order moment of the function $H$, which 
has the value $2/\sqrt{3}$, while $\beta_1$ approaches 0. To our knowledge, expression (106) for the grayness factor was established for the first time by King \cite{king1963}, who 
called it  the ``greenhouse factor''. We point out that in his derivation the factor $\mu_0$ is missing in both the numerator and 
denominator of the left-hand side of his Eq. (25), two omissions that happily cancel each other.\\
As to the temperature distribution of a semi-infinite atmosphere in local thermodynamic equilibrium, it is readily found by simply identifying the source function 
$S(\tau,\mu_0)$ in Eq. (100) with the spectrally integrated formula of Planck, i.e., $(\sigma/\pi)T^4(\tau,\mu_0)$. Remembering the relation $\sigma T_0^4=F_0$ defining the temperature $T_0$, 
we arrive at the simple expression
\begin{equation}
\label{eq107}
\frac{T(\tau,\mu_{0})}{T_{0}}= \left[\frac{1}{4}B(\tau,\mu_0)\right]^{1/4},
\end{equation}
which is exactly the result that we would get if we let $\tau_b \to \infty$ in (79). The graphical representation of this temperature distribution is very 
much like that computed for a finite medium with $\tau_b=10$ (Fig. 1). As $B(0,\mu_0) = H(\mu_0)$ and $B(\infty,\mu_0) = \sqrt{3}\mu_0 H(\mu_0)$, the 
extreme temperature values are seen to be given by
\begin{equation}
\label{eq108}
\frac{T(0,\mu_{0})}{T_0}=\left[\frac{H(\mu_0)}{4}\right]^{1/4} \quad,\quad\frac{T(\infty,\mu_{0})}{T_0}=\left[\frac{\sqrt{3}}{4}\mu_0 H(\mu_0)\right]^{1/4},
\end{equation}  
from which follows the exact relationship $T(\infty,\mu_{0})/T(0,\mu_{0})=(\sqrt{3}\mu_0)^{1/4}$.\\
Equation (107) can be evaluated for grazing incidence of the incoming rays, reminding ourselves that by 
virtue of the relations (95)-(97), the limit of $B(\tau,\mu_0)$ for $\mu_0\rightarrow 0^+$ is $H(0)=1$ if 
$\tau=0$, and 0 if $\tau>0$. For grazing incidence, the temperature of the atmosphere is therefore zero everywhere except in the ``layer'' at $\tau=0$, where it 
takes the value $T_0/\sqrt{2}$.\\
Now, for $\mu_0>0$, we may calculate the derivative of the temperature distribution by substituting the expression (97) for $B(\tau,\mu_0)$ 
into (107), then appealing to Eq. (23) in \cite{bergeatandrutily1998}, viz.
\begin{equation}
\label{eq109}
\mu_0\frac{\partial \eta}{\partial \tau}(\tau,\mu_0)=\phi(\tau)-\eta(\tau, \mu_0) \quad(\tau>0),
\end{equation}  
in order to get
\begin{equation}
\label{eq110}
4\left [\frac{T(\tau,\mu_{0})}{T_{0}}\right ]^3\frac{T'(\tau,\mu_{0})}{T_{0}}=\frac{1}{4}H(\mu_0)[\phi(\tau)-\eta(\tau, \mu_0)].
\end{equation}
The sign of $T'(\tau,\mu_{0})$ is therefore the same as that of the function $\phi(\tau)-\eta(\tau,\mu_0)$, that we have tabulated. 
We arrive at the following conclusion:\\
- for $0<\mu_0<1$, the temperature has a maximum in the interval $]0,+\infty[$, achieved at a depth $\tau$ which is closer to 0 the smaller 
$\mu_0$, and then approaches $T(\infty,\mu_{0})=[\frac{\sqrt{3}}{4}\mu_0 H(\mu_0)]^{1/4}T_0$,\\
- for $\mu_0=1$, the temperature increases with $\tau$ up to the value $T(\infty,1)=[\frac{\sqrt{3}}{4}H(1)]^{1/4}T_0$.\\

\subsection{Semi-infinite atmospheres with sources at infinity}
In the second category of semi-infinite models, suiting the giant planets, we choose a vanishing incident radiation as the upper boundary condition. We thus let\\
$(iii)$ the incident radiation be naught at the top of the atmosphere
\begin{equation}
\label{eq111}
I(0,\mu)=0\quad(-1\leq\mu<0),
\end{equation}%
and choose ``stellar-type'' conditions at infinity:\\
$(v)$ in the deep layers of the atmosphere, the radiance $I(\tau,\mu)$ increases more slowly than $\exp(\tau/\mu)$ for any $\mu>0$:
\begin{equation}
\label{eq112}
\lim_{\tau\to \infty}I(\tau,\mu)\exp(-\tau/\mu)=0\qquad (0<\mu\leq 1).
\end{equation}
Thus we retrieve the model of a gray stellar atmosphere in radiative equilibrium that was studied by Hopf \cite{hopf1934}, Chandrasekhar \cite{chandrasekhar1950} 
and others. The formal solution of the transfer equation reduces to the integrals in (11) and (12), and hence the functions $J_0$, $F_0^{\uparrow}$ and $F_0^{\downarrow}$ 
on the right-hand side of Eqs. (13)-(15) vanish. We look for a solution, unbounded at infinity, of the {\it homogeneous} integral equation (20) with 
$\tau_b=\infty$. Hopf \cite{hopf1934} showed that an infinite number of such solutions exist, all defined up to a multiplicative constant, but 
that only one corresponds to a given emergent flux $F$, viz.
\begin{equation}
\label{eq113}
S(\tau)=\frac{3F}{4\pi}[\tau+q(\tau)].
\end{equation}
Here, $q(\tau)$ denotes Hopf's function, which may be expressed as \cite{ivanov1973}
\begin{equation}
\label{eq114}
q(\tau)=\frac{1}{\sqrt{3}}\left \lbrace 1+\frac{1}{2}\int_{0}^{1}\frac{g(v)}{H(v)}[1-\exp(-\tau/v)]\mathrm{d}v\right \rbrace.
\end{equation}
One may then deduce the radiance at every point in the atmosphere, in much the same way as in Subsection 8.2. We get
\begin{equation}
\label{eq115}
I(\tau,\mu)=\frac{3F}{4 \pi}\left [\tau+q(\tau)+\frac{1}{\sqrt{3}}\mu \eta(\tau,-\mu)\right ].
\end{equation} 
The resulting radiance has been published often in the literature, but to our knowledge in forms more intricate than (115). At the surface of the atmosphere, 
Hopf's function has the value $q(0) = 1/\sqrt{3}$, while $\eta(0,\mu)$, given by (95), at once shows that $I(0,\mu) = 0$ 
for $\mu<0$. On the other hand, for $\mu>0$, we retrieve Hopf's result \cite{hopf1934}
\begin{equation}
\label{eq116}
I(0,\mu)=\frac{\sqrt{3}F}{4\pi}H(\mu).
\end{equation}
As $\tau\to\infty$, Hopf's function remains bounded, approaching $q(\infty)\sim 0.710$, while the function $\eta$ tends to $\sqrt{3}$, so that
\begin{equation}
\label{eq117}
I(\tau,\mu)\sim \frac{3F}{4 \pi}[\tau+q(\infty)+\mu]\sim \frac{3F}{4 \pi}\tau\quad(\tau\rightarrow\infty).
\end{equation}
The radiance diverges at infinity, but without violating the condition (112).\\
The upward and downward fluxes may be deduced from expression (115) for the radiance, much as in Subsection 8.2. We get
\begin{equation}
\label{eq118}
F^{\uparrow}(\tau)= \frac{3 F}{4}[\tau+q(\tau)]+\frac{\sqrt{3}}{2}F\int_{-1}^{0}\eta(\tau,\mu)\mu^2\mathrm{d}\mu
\end{equation}
\begin{equation}
\label{eq119}
F^{\downarrow}(\tau)= \frac{3 F}{4}[\tau+q(\tau)]-\frac{\sqrt{3}}{2}F\int_{0}^{1}\eta(\tau,\mu)\mu^2\mathrm{d}\mu
\end{equation}
Note that the difference of these two fluxes just gives $F$, in agreement with (8), because, as may be shown,  
$\int_{-1}^{+1}\eta(\tau,\mu)\mu^2\mathrm{d}\mu = 2/\sqrt{3}$.\\ 
The calculation of the temperature distribution in an atmosphere in local thermodynamic equilibrium, with the source function given by (113), is 
straightforward; it leads to the famous relation of Hopf \cite{hopf1934}
\begin{equation}
\label{eq120}
\frac{T(\tau)}{T_{ef}}= \left\lbrace\frac{3}{4}[\tau+q(\tau)]\right\rbrace^{1/4},
\end{equation}
in which the effective temperature is defined by $\sigma T_{ef}^4=F$. 

\section{Conclusion}
The generalized problem we have outlined originated with two prototype problems of the classical theory of planetary atmospheres. The first one
 is of relevance to situations where shortwave radiation interacts with the planet's atmosphere, modeled as a slab under parallel-beam illumination from the sun 
and overlying a partly reflecting ground. In the second problem, there is no incident radiation on the upper surface of the atmosphere but only on the lower 
one, issuing from an isotropically emitting ground, a model appropriate to radiation transfer through a planetary atmosphere in the thermal (or infrared) regime. 
In a strictly gray model, in which both the sun's radiation and the planet's ground emission are scattered and absorbed by the planet's atmosphere, the combined 
boundary conditions prevail simultaneously, providing us with the model that we offer in mathematical elaboration in the present article.\\
Of course, this physically rather unsophisticated model is not realistic, but in exchange it is susceptible to analytical tractability. It yields 
the radiance field in a semi-infinite atmosphere illuminated from the outside or heated by sources at infinity, for which the analytical developement could be pushed 
as far as it goes. In the case of an optically finite atmosphere, we were able to calculate exactly the source function for each depth, as well as the 
upward and downward fluxes across the boundaries. As a result, it was possible to determine the temperature of the atmosphere as well as that of the ground.\\
Models more realistic than the one here presented may inspire further developments and extensions. In particular, we think of the possibility of applying each set 
of boundary conditions to two separate frequency domains in such a manner that the radiation interactions remain coupled by the radiative equilibrium constraint. 
Models of that kind were proposed early, notably by Emden \cite{emden1913}, and Milne \cite{milne1922}, who realized that the angular variable $\mu_0$ of the gray models 
could be extended to the range $]1,+\infty[$ in what is sometimes termed semi-gray models. This opens up the possibility of applying the results for gray 
media to those of semi-gray media, which represents a first step towards more realistic models.\\
Semi-gray models are thus seen to constitute the natural next step in a hierarchy of analytical models of radiative transfer within planetary atmospheres. We 
intend to propose such a semi-gray model in two companion papers. The present study can therefore be judged as an intermediate step in the application of 
the exact solution to the thermal radiation problem, as given recently in \cite{rutilyetal2004}, and the forthcoming semi-gray model.

\appendix
\renewcommand{\theequation}{\Alph{section}\arabic{equation}}
\setcounter{equation}{0}
\setcounter{section}{1}

\section*{Appendix. Analytical expressions for the $B$- and $\xi_0$-functions in a conservative plane-parallel medium.}
The expressions below involve the auxiliary functions $T$, $g$ and $H$ defined by Eqs. (88), (90) and (86), respectively. 
We will also need the two basic functions $\zeta_{\pm}=\zeta_{\pm}(\mu)$ already introduced for a finite slab in Section 6 of \cite{chevallierandrutily2005}, 
as well as the moments of order 0 and 1 of the $X$- and $Y$-functions defined by (26). These moments are tabulated in \cite{vandehulst1980}, for instance.\\
For $\tau_b<\infty$ and $\mu>0$, we have 
\begin{equation}
\label{eqA1}
B(\tau,\mu)=c(\mu)\exp(-\tau/\mu)+\frac{1}{2}H(\mu)\mu[\zeta_{-}(-\mu)\mathfrak{F}_+(\tau,\mu)+\zeta_{+}(-\mu)\mathfrak{F}_-(\tau,\mu)],
\end{equation}
where $c(\mu)$ is given by (92) and 
\begin{eqnarray}
\label{eqA2}
\lefteqn{\mathfrak{F}_{+}(\tau,\mu)=\frac{3}{2}(\tau_b-2\tau+2\mu)\beta_0\nonumber}\\
&&\qquad\quad+\frac{1}{2}\int_0^1\frac{g(v)}{H(v)}\,\zeta_{+}(v)[\frac{\exp(-\tau/v)}{v-\mu}-\frac{\exp(-(\tau_b-\tau)/v)}{v+\mu}]\mathrm{d}v,
\end{eqnarray}
\begin{eqnarray}
\label{eqA3}
\lefteqn{\mathfrak{F}_{-}(\tau,\mu)=\frac{3}{2}(\alpha_1+\beta_1)\nonumber}\\
&&\qquad\quad+\frac{1}{2}\int_0^1\frac{g(v)}{H(v)}\,\zeta_{-}(v)[\frac{\exp(-\tau/v)}{v-\mu}+\frac{\exp(-(\tau_b-\tau)/v)}{v+\mu}]\mathrm{d}v.
\end{eqnarray}
These expressions simplify in a semi-infinite space, since then the functions $\mathfrak{F}_{\pm}$ coincide with the $\mathfrak{F}$-function defined 
by (93). Likewise, we have $\zeta_{\pm}=1$ in a half-space, and Eq. (A1) reduces to Eq. (97).\\
These formulae are new, they are given here without any proof. We obtained them by solving an integral equation of Cauchy type, satisfied by the function $B$, along the general lines of \cite{rutilyetal2007}. Our equation (A1) is a particular case of the general equations in Section IV of \cite{rutilyetal2007}, from which it follows by choosing $a=1$ and $c_0(z)=\exp(-\tau/z)$.\\
The expression of the $\xi_0$-function has already been given in \cite{rutilyetal2004}, Eqs. (62)-(63), which we repeat here for convenience. For $\tau_b<\infty$
\begin{equation}
\label{eqA4}
\xi_0(\tau)=\frac{1}{2}[1+ \beta_0 F_{-}(\tau)],
\end{equation}
where
\begin{eqnarray}
\label{eqA5}
\lefteqn{F_{-}(\tau)=\frac{3}{4}(\tau_b-2\tau)(\alpha_1+\beta_1)\nonumber}\\
&&\qquad+\frac{1}{2}\int_0^1\frac{g(v)}{H(v)}\zeta_{-}(v)[\exp(-\tau/v)-\exp(-(\tau_b-\tau)/v)] \d v,
\end{eqnarray}
while for $\tau_b=\infty$, we have $\xi_0(\tau)=1$ as expected.

\section*{Acknowledgements}
The authors wish to acknowledge Dr. O. Titaud (Centro de Modelamiento Matem\'atico, Santiago, Chile) for helpful comments on a previous version of the manuscript.

\end{document}